\documentclass[aps,twocolumn,superscriptaddress]{revtex4-1}

\usepackage{graphicx}
\usepackage{color}
\usepackage{graphicx}
\usepackage{amsmath}
\usepackage{dcolumn}

\usepackage{epsfig}
\usepackage{graphics}
\usepackage{latexsym}
\usepackage{amsfonts}
\usepackage{amssymb}

\usepackage{bbm}
\usepackage{bbold}
\usepackage{bm}   

\usepackage{mathtools}
\usepackage{booktabs}
\usepackage{placeins} 

\usepackage[T1]{fontenc}
\usepackage[utf8]{inputenc} 

\usepackage{textgreek} 

\DeclareMathOperator{\Arg}{Arg}
\DeclareMathOperator{\Log}{Log}
\DeclareMathOperator{\sign}{sign}

\newcommand{\am}{\mathcal{A}}
\newcommand{\bet}{\eps}

\newcommand{\Ca}{\text{Ca}}
\newcommand{\cinf}{\bar{c}}
\newcommand{\Da}{\text{Da}}
\newcommand{\Deff}{D_\mathrm{eff}}
\newcommand{\Ds}{D_\mathrm{s}}

\newcommand{\dif}[1]{\mathrm{d}#1}
\renewcommand{\d}{\text{d}}
\newcommand{\eqdef}{\equiv}
\newcommand{\eps}{\epsilon}
\newcommand{\fo}{f_\mathrm{o}}
\newcommand{\fou}[1]{\hat{#1}}
\newcommand{\fouop}{\mathcal{F}}
\newcommand{\fouopbra}[1]{\fouop[#1]}
\newcommand{\Gam}{\Gamma}
\newcommand{\Gamo}{\Gamma_\mathrm{o}}
\newcommand{\gam}{\gamma}
\newcommand{\gamo}{\gam_\mathrm{o}}
\newcommand{\hanop}{H}

\newcommand{\hi} [1]{\breve{#1}}  
\newcommand{\hilop}{\mathcal{H}}
\newcommand{\hilopbra}[1]{\hilop[#1]}
\newcommand{\hifo}{\hi{f}_\mathrm{o}}
\newcommand{\ie}{i.e.}

\newcommand{\Imag}{\mathrm{Im}}
\newcommand{\indcase}{\chi}
\newcommand{\intii}{\int_{-\infty}^{\infty}}
\newcommand{\kap}{\kappa}

\newcommand{\nuexp}{\zeta}
\newcommand{\Pe}{\text{Pe}}
\newcommand{\pos}[1]{\left[ #1  \right]_{+}}
\newcommand{\Psidag}{\Psi^\dagger}
\newcommand{\Psio}{\Psi_\mathrm{o}}
\newcommand{\qflux}{\mathcal{Q}}
\newcommand{\PV}{\mbox{p.v.}}
\newcommand{\Renum}{\text{Re}}
\newcommand{\Real}{\mathrm{Re}}
\newcommand{\riz}[1]{\check{#1}}  
\newcommand{\rizop}{\mathcal{R}}
\newcommand{\rizopbra}[1]{\rizop[#1]}

\newcommand{\taueff}{\tau_\mathrm{eff}}
\newcommand{\taur}{\tau_\mathrm{r}}
\newcommand{\tsing}{{t^\star}}
\newcommand{\us}{x^*}
\newcommand{\usf}{u_\mathrm{s}}
\newcommand{\Uo}{U}
\newcommand{\xvmax}{x_\mathrm{vmax}}
\newcommand{\xsing}{{x^\star}}
\newcommand{\zcom}{\mathrm{z}}

\newcommand{\vc}[1]{\mathbf{#1}}
\newcommand{\vq}{\vc{q}}

\newcommand{\vu}{\vc{u}}
\newcommand{\vus}{\vu_\mathrm{s}}

\newcommand{\vx}{\vc{x}}

\usepackage{color}
\usepackage[dvipsnames]{xcolor}

\begin{document}

\title{Exact solutions for viscous Marangoni spreading}

\author{Thomas Bickel}
\affiliation{Univ. Bordeaux, CNRS, Laboratoire Ondes et Mati\`ere d'Aquitaine, F-33400 Talence, France}
\author{Fran\c cois Detcheverry}
\affiliation{University of Lyon, Universit\'e Claude Bernard Lyon 1, CNRS, Institut Lumi\`ere Mati\`ere, F-69622 Villeurbanne, France}

\date{\today}
\begin{abstract}
When surface-active molecules are released at a liquid interface, 
their  spreading dynamics is controlled by Marangoni flows. 
Though such  Marangoni spreading was investigated in different limits, 
exact solutions  remain very few. 
Here we consider the spreading of an insoluble surfactant along the interface of a deep fluid layer. 
For two-dimensional Stokes flows,
it was recently shown that the non-linear transport problem 
can be exactly mapped to a complex Burgers equation  [Crowdy, {\it SIAM J. Appl. Math.}, {\bf 81}, 2526 (2021)].
We first present a very simple derivation of this equation.   
We then provide  fully explicit solutions 
and find that varying the initial surfactant distribution  -- pulse, hole, or periodic --
results in distinct spreading behaviors. 
By obtaining the fundamental solution, we also discuss the influence of surface diffusion.   
We identify  situations where spreading can be described as an effective diffusion process
but observe that this approximation is not generally valid.  
Finally, the case of  a three-dimensional flow with axial symmetry is briefly considered. 
Our findings should provide reference solutions for Marangoni spreading, 
that may be tested experimentally  with fluorescent or photoswitchable surfactants.
\end{abstract}

\maketitle

\section{Introduction}

The spreading of surface-active species at aqueous interfaces 
is a long-standing issue in interfacial science~\cite{Matar_sm-2009}. 
Surfactants, of which amphiphilic molecules are the prime example, 
accumulate at interfaces where they lower the surface tension~\cite{book_dGbwq-CapWet}. 
As a consequence, 
a gradient in surface concentration induces a Marangoni stress at the interface~\cite{Scriven_nat-1960,Manikantan_jfm-2020}. 
Inhomogeneities in the distribution of surfactants thus drive a liquid flow, 
that in turn couples with the surfactant distribution. 
This complex feedback mechanism may eventually lead to the rigidification of the interface 
that can alter the rising motion of  gas bubbles~\cite{BelFdhila_pf-1996,Takagi_arfm-2011,Palaparthi_jfm-2006,Crowdy_jem-2021}. 
Another striking effect observed in microfluidic experiments 
is that traces of surfactants can severely limit the drag reduction of superhydrophobic surfaces~\cite{Peaudecerf_pnas-2017,Landel_jfm-2020,Baier_jfm-2021}. 
More generally, 
Marangoni stresses due to the presence of surfactants, even at a very low concentration, 
are ubiquitous  
through fundamental processes~\cite{Bickel_epje-2019,Bickel_prf-2019,Venerus_sr-2015}, 
the natural world~\cite{Trinschek_sm-2018,Stetten_cocis-2018,Botte_jms-2005} 
and industrial applications~\cite{Morciano_ees-2020}. 
The consequences are essential for a variety of phenomena,  
including film thickness in coating ~\cite{Quere_arfm-1999},  
dispersion relation for capillary waves~\cite{Sauleda_jcis-2022} 
and stability  of foams~\cite{book_Cantat-Foams,Breward_jfm-2002}. 

The dynamics of surfactant spreading is also relevant in the field of active matter 
through the propulsion mechanism of Marangoni swimmers~\cite{Fei_cocis-2017,Grosjean_epje-2018,book_nplks-SelfOrganizedMotion}. 
When a particle filled with surface-active molecules such as camphor  
is placed at the air-water interface, 
the resulting surface-tension gradient drives the spontaneous motion of the particle, 
even if the latter is perfectly symmetric~\cite{Rednikov_jcis-1994,Boniface_pre-2019,Ender_epje-2021}.  
Although camphor swimmers are known for centuries~\cite{Boniface_pre-2019}, 
their rich individual and collective dynamics  is still a matter of investigation~\cite{Suematsu_cej-2018,Nakata_pccp-2015,Dietrich_prl-2020,Boniface_prf-2021}.  
To elucidate their propulsion mechanism, 
a fine understanding of the transport of surfactants is required. 
In particular, it has been suggested 
both experimentally~\cite{Suematsu_langmuir-2014} and  theoretically~\cite{Kitahata_jcp-2018,Bickel_sm-2019}
that the transient spreading of camphor molecules 
could be described by an effective diffusion process that would account for the advection by the Marangoni flow. 
This point is actually quite subtle and requires an in-depth investigation.

When a surfactant is released at the air-liquid interface, 
the surfactant-covered domain grows in time and 
its size $L(t)$ is governed by a balance between viscous stresses and surface tension gradients. 
At large time, the size generally follows the scaling law  $L(t) \sim t^{\alpha}$. 
The exponent~$\alpha$ depends on the dominant features of the system such as inertia, gravity, capillarity or viscous dissipation, among others. 
Although most of previous studies focused on thin films~\cite{Jensen_jfm-1992,AfsarSiddiqui_acis-2003,Dussaud_jfm-2005,Craster_rmp-2009,Swanson_jem-2015}, where the lubrication approximation applies, we consider here the deep layer limit.  
Regarding the dynamics of surfactant spreading, 
asymptotic self-similar solutions have been thoroughly discussed in the wake of the seminal work of Jensen~\cite{Jensen_jfm-1995}. 
More recently, 
the hydrodynamic signature of surfactant transport in a semi-infinite liquid 
has also been investigated at steady state~\cite{Roche_prl-2014,LeRoux_pre-2016,Bandi_prl-2017,Mandre_jfm-2017}. 
This line of works focuses mainly on the large Reynolds limit, 
where the surfactant-driven flow develops in a boundary layer.

In contrast, 
much less is known regarding the  properties of Marangoni spreading with Stokes flow.
In this case, 
the vorticity created by the Marangoni shear stress at the interface penetrates deep into the liquid, so that the  transport equations  are generally both nonlocal and nonlinear~\cite{Matar_sm-2009}. 
An important step forward was achieved, however, when it was recognized that the interfacial velocity can be expressed as a convolution of the surface concentration~\cite{Thess_prl-1995}. 
Another essential finding, achieved recently by Crowdy~\cite{Crowdy_siamjam-2021},  
has revealed that the  mathematical problem for insoluble surfactant can be mapped to a complex Burgers equation,
opening new perspectives for its resolution.   

The aim of the present work is to exploit those advances 
to build a complete set of exact solutions for viscous Marangoni spreading.   
In contrast to self-similar solutions that apply asymptotically, 
our  solutions are valid at all time.  
We find that the non-linear character of the spreading problem  
leads to a rich variety of possible behaviors  
and that the initial surfactant distribution has a key influence on the subsequent evolution. 
Besides,   
our exact solutions provide  reference cases and a range of physical insights. 
In particular,   
the question of  Marangoni spreading as an effective diffusion process can now be  settled. 

The study is organized as follows. 
In Sec.~\ref{sec:model}, 
we first present the hydrodynamic model with its underlying assumptions and 
propose a particularly simple derivation of the complex Burgers equation 
that governs the  dynamics of insoluble surfactants  in one dimension. 
Section~\ref{sec:1d} focuses on transient Marangoni spreading  in the absence of diffusion.   
Exact solutions  are obtained for a variety of initial surfactant distributions. 
The influence of surface diffusion is investigated in Sec.~\ref{sec:1d_diff}, 
where we provide a fundamental solution and discuss the resulting effective diffusion coefficient. 
Finally, 
we examine in Sec.~\ref{sec:2d} the spreading dynamics in two dimensions with axial symmetry.  
We conclude in Sec.~\ref{sec:dis} with a physical discussion of our findings.

\section{From Marangoni spreading  to  Burgers equation}
\label{sec:model}

\subsection{Dimensionless numbers and simplifying assumptions}

\begin{figure}[t!]
\hspace*{-2mm}
\includegraphics[width=9cm]{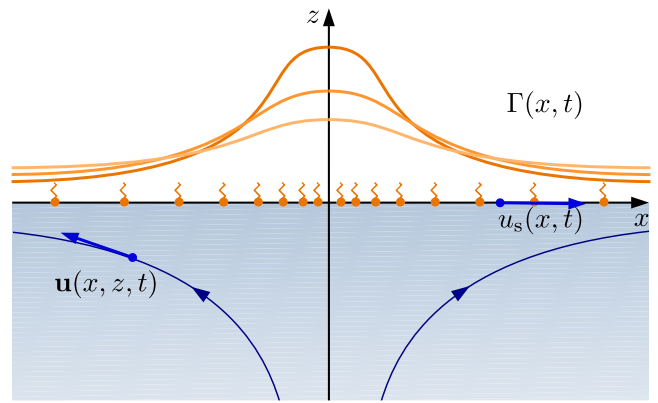}
\caption{
Spreading of an insoluble surfactant above a semi-infinite liquid layer with two-dimensional flow. 
The inhomogeneous distribution of surfactant~$\Gamma(x,t)$ 
induces a Marangoni flow 
with bulk velocity~$\vu(x,z,t)$ and interfacial velocity~$\usf(x,t)$.
}  
\label{fig:schema}
\end{figure}

We consider the spreading of surfactants at the air-water interface, as illustrated in Fig.~\ref{fig:schema}.
Initial inhomogeneities in the surfactant  distribution 
induce Marangoni constraints and a fluid flow that drives the system toward a homogeneous state.  
Our goal is to provide an exact description of this relaxation process. 
We assume a two-dimensional flow  and an infinitely deep liquid layer. 
The liquid is Newtonian and incompressible, 
with dynamic viscosity~$\eta$, mass density~$\rho$ and kinematic viscosity~$\nu=\eta/ \rho$.
The surface-active molecules are insoluble, \ie{} they are irreversibly adsorbed  at the interface. 
We note~$\Ds$ their surface diffusion coefficient.  

To introduce further modeling assumptions, we first discuss the relevant dimensionless numbers. 
The contribution of advection to momentum and mass transport can be rationalized by the Reynolds and P\'eclet numbers
\begin{align}
    \Renum \eqdef \frac{L \Uo}{\nu}   \quad \text{and} \quad   \Pe \eqdef \frac{L \Uo}{\Ds},
\end{align}
with~$L$ the length scale of the initial concentration perturbation  
and~$\Uo$ the velocity associated with the corresponding Marangoni flow. 
Both quantities may span a large range in experiments,  
namely $L=10^{-3}-10^{-1}\,$m and $\Uo=10^{-3}-1\,$m$\,$s$^{-1}$, 
resulting in Reynolds number $\Renum=1-10^5$ much above or around unity.  
Because the surface diffusion coefficient~$\Ds \simeq 10^{-9}\,$~m$^2\,$s$^{-1}$~\cite{Shmyrov_langmuir-2019}
is several orders of magnitude smaller than the momentum transport 
coefficient~$\nu \simeq 10^{-6}\,$~m$^2\,$s$^{-1}$, 
the hierarchy~$\Renum \ll \Pe$ always applies~\cite{Thess_jfm-1997}. 
The  thickness of the hydrodynamic boundary layer is therefore much larger than that of the mass boundary layer~\cite{LeRoux_pre-2016}. 
In the following, we neglect fluid inertia and focus on the Stokes regime. 
This assumption is reasonable when $L$ lies below or within  the millimeter range  
and~$U$ in the mm$\,$s$^{-1}$ range at most. 
While this choice of the Stokes limit  is certainly restrictive, 
we will show in the following that it is interesting in its own right. 
Emphasis is placed on nonlinearities that occur in the mass transport equation 
and on the methods to handle them exactly.

The spreading of surfactant can  induce an unsteady displacement of the free interface. 
The competition between surface deformation and  viscous stress is quantified by the capillary number
\begin{align}
    \Ca \eqdef \frac{\eta \Uo}{\gamma},  \label{eq:Ca}
\end{align}
with $\gamma$ the surface tension of the interface.  
With $\eta \simeq 10^{-3}\,$Pa$\,$s, 
$\gamma = 10^{-2}-10^{-1}\,$N$\,$m$^{-1}$ 
and the range of velocity scale given above,  
one gets $\Ca = 10^{-5}-10^{-1}$. 
We can therefore neglect capillary effects 
and assume that the interface remains flat in the formulation of the hydrodynamic problem. 
Still, interfacial deformations can be determined afterward by invoking the normal stress continuity condition~\cite{Jensen_jfm-1995}.  

Finally, some surface-active species such as camphor or alcohol molecules
may evaporate from the liquid to the gas phase. 
To account for this phenomenon, we include a first-order kinetic evaporation with rate~$k$. 
The ratio of evaporative to convective mass transport  is set by the Damk\"{o}hler number
\begin{align}
    \Da \eqdef \frac{kL}{\Uo}, 
\end{align}
that can take values in a very wide range, spanning from negligible  ($\Da \ll  1$) to prevalent  ($\Da \gg  1$) evaporation.

\subsection{Marangoni flow and surfactant transport}

In the viscous regime, the velocity field~$\vu=(u,v)$ and pressure~$p$ satisfy the incompressible Stokes equations
\begin{align}
   \bm{\nabla} \cdot \vu &= 0, \qquad    \eta \Delta   \vu = \bm{\nabla} p, 
   \label{eq:stokes}
\end{align} 
with $\bm{\nabla}$ the nabla operator and $\Delta$ the Laplacian. 
The velocity is assumed to vanish far away from the initial disturbance.
Along the free interface, 
variations in surface tension~$\gam$ induce a tangential stress given by the Marangoni boundary condition
\begin{align}
    \eta \,\partial_z u \big\vert_{z=0} = - \partial_x \gam, 
    \label{eq:marangoniBC}
\end{align}
where the interface  remains flat and corresponds to $z=0$. 
The surface tension  is related to the local surfactant concentration $\Gam$ 
through an equation of state  $\gam=\gamo - \kappa \Gamma$ 
that is linear~\cite{book_birdi-HandSurfaceColloidChem_chapter4},  
a convenient but approximate assumption~\cite{Pawar_pf-1996,Swanson_jem-2015}.  
Surface activity is then measured by the positive constant~
$\kap \eqdef -  \partial \gam / \partial \Gam$.  
Finally, 
the  concentration~$\Gam(x,t)$ of surfactant evolves according to the advection-diffusion equation
\begin{align}
  \partial_t \Gam + \partial_x \left( \Gam \usf \right) = \Ds \, \partial^2_{xx} \Gam - k \Gam,   
  \label{eq:advdiff}
\end{align}
where $\usf(x,t) \eqdef u(x,z=0,t)$ is the velocity at the interface. 
The last term in Eq.~\eqref{eq:advdiff} accounts for evaporation.

\subsection{Closure relation and Thess' equation}

Since the Stokes equation is linear, 
it should be possible to relate the flow anywhere in the bulk 
to the distribution of Marangoni stress at the interface, and thus ultimately to the distribution of surfactant. 
This step has indeed been achieved by Thess and collaborators~\cite{Thess_prl-1995,Thess_ps-1996,Thess_jfm-1997} (see also Ref.~\cite{Pismen_prl-1997}). 
The end result is  a closure relation giving the surface velocity as a convolution of surfactant concentration 
with a dimension-dependent kernel. 
When the surface velocity is one-dimensional, 
the closure relation reads 
\begin{equation}
    \usf(x,t) = \frac{\kap}{2\eta} \, \hilop[\Gam(x,t)].
    \label{eq:closure1D}
\end{equation}
Here, we define the Hilbert transform~$\hilop$ of a function~$f$ 
\begin{equation}
    \hilopbra{f}(x) \eqdef \hi{f}(x) \eqdef \frac{1}{\pi} \PV \int_{-\infty}^{\infty} \frac{f(x')}{x-x'} \d x', 
    \label{eq:hilbert}
\end{equation}
where  the integral is  understood in the sense of Cauchy principal value~($\PV$).
The closure relation of Eq.~\eqref{eq:closure1D} is non-local: 
the velocity at position~$x$ depends on the surfactant distribution everywhere on the surface. 
This is in clear contrast with the thin layer limit, 
where the surface velocity is simply proportional to the concentration 
gradient~$\partial_x \Gam$~\cite{Thess_jfm-1997}.

It is convenient at this point to switch to dimensionless variables. 
From now on, we focus on 
the reduced surfactant concentration $f=\Gamma/\Gamo$ 
and the reduced velocity $u=\usf/\Uo$, 
with $\Gamo$ the concentration scale 
and $\Uo= \kap \Gamo/2\eta$ the characteristic velocity associated with the Marangoni flow. 
The length and time scales are also expressed in units of $L$ and $\tau=L/\Uo$, 
with~$L$ the relevant size pertaining to the initial perturbation. 
Expecting no confusion by the reader, 
we keep the same notations~$x$ and~$t$ for the dimensionless variables. 
According to Eq.~\eqref{eq:closure1D}, 
the closure relation in dimensionless form simply reads  $u(x,t)=\hi{f}(x,t)$. 

Putting all pieces together, 
the  flow and surfactant dynamics described by  Eqs.~\eqref{eq:stokes}--\eqref{eq:advdiff} 
can finally be replaced by a unique equation originally introduced by Thess~\cite{Thess_ps-1996}
\begin{equation}
  \partial_t f + \partial_x ( f \hi{f} ) = - \alpha f + \bet \, \partial^2_{xx} f  \ ,  
  \label{eq:thesseq}
\end{equation}
where $\alpha$  and $\bet$  stand respectively for the Damk\"{o}hler number $\alpha \eqdef \Da$ 
and the inverse P\'eclet numer $\bet \eqdef \Pe^{-1}$. 
By ``integrating out'' the  features of the flow, 
the coupled problem of momentum and mass transport is  reduced to a single equation, 
which is, however, nonlinear and nonlocal. 

\subsection{Complex Burgers equation}

Although some specific solutions of Thess' equation 
have been discussed previously in the literature~\cite{Thess_ps-1996,Thess_jfm-1997,Pismen_prl-1997,Bickel_sm-2019}, 
a systematic method for solving the problem is still missing. 
A major breakthrough has been achieved very recently by Crowdy~\cite{Crowdy_siamjam-2021}, 
recognizing that the problem defined by Eqs.~\eqref{eq:stokes}-\eqref{eq:advdiff} 
can be mapped to a complex Burgers equation. 
The derivation is based on the complex variable representation 
of the stream function and therefore relies on advanced properties of analytic functions.
Here we propose a different route, 
inspired by an approach originally developed in the context of vorticity transport~\cite{Constantin_cpam-1985}.

By exploiting properties of Hilbert transforms,  
let us show that Crowdy's result can be derived in a straightforward manner. 
The first step is to apply the Hilbert operator to Eq.~\eqref{eq:thesseq} to obtain 
\begin{align}
  \partial_t \hi{f} +   \hi{f} \partial_x  \hi{f} - f \partial_x f  = - \alpha \hi{f} + \bet  \, \partial^2_{xx} \hi{f}.
  \label{eq:hilbertthesseq}
\end{align}
Here, we made use of the  relations 
$\hilopbra{\partial_x f} = \partial_x \hi{f}$, 
$\hilopbra{f g}= \hi{f} g + f \hi{g} + \hilopbra{\hi{f} \hi{g}}$
and $\hilop^2[f]=-f$, from which one also gets $\hilopbra{ f \hi{f} }= (\hi{f}^2-f^2)/2$~\cite{book_King-HilbertTransformII,Constantin_cpam-1985}. 
The second step  is to introduce the complex function $\Psi(x,t)$ as
\begin{align}
 \Psi(x,t) \eqdef \hi{f}(x,t) - if(x,t).    \label{eq:Pisdefinition}
\end{align}
Combining Eqs.~\eqref{eq:thesseq} and~\eqref{eq:hilbertthesseq}, 
it is then a simple calculation to recover Crowdy's equation~\cite{Crowdy_siamjam-2021} 
\begin{align}
  \partial_t \Psi + \Psi \partial_x  \Psi= - \alpha \Psi + \bet  \, \partial^2_{xx} \Psi, 
  \label{eq:burgers}
\end{align}
with an additional term that accounts for evaporation. 
This equation expresses that in the present setting,  
Marangoni spreading is  formally governed by a viscous Burgers equation for the complex function $\Psi(x,t)$. 
If a solution can be found,  
the  interfacial velocity and the concentration are deduced from  
\begin{align}
u(x,t)=\Real [\Psi(x,t) ], \qquad f(x,t) = - \Imag [\Psi(x,t)].
\end{align}
Note that, since $f$ is a concentration, it should always satisfy $f(x,t) \geq 0$. 

The inverse P\'eclet number $\bet$ appearing in Eq.~\eqref{eq:burgers} 
plays the role of viscosity in the more common form of the  Burgers equation, 
which was initially introduced in the context of 
fluid turbulence~\cite{book_Burgers-NonlinDifEq,Bonkile_pramanajp-2018}. 
The effect of evaporation is embodied through an extra linear term monitored by the Damk\"{o}hler number $\alpha$. 
The main accomplishment of shifting from Thess' to Crowdy's formulation 
is that locality is now restored in the  Burgers~Eq.~\eqref{eq:burgers}. 
Moreover, a rigorous mathematical transformation allows 
to further transform the nonlinear Burgers equation into a linear partial differential equation~\cite{Hopf_cpam-1950}. 
As a consequence, analytical solutions for the transient spreading of insoluble surfactants 
can be derived and discussed in a systematic way.
We now proceed to do so. 

\begin{figure*}[t!]
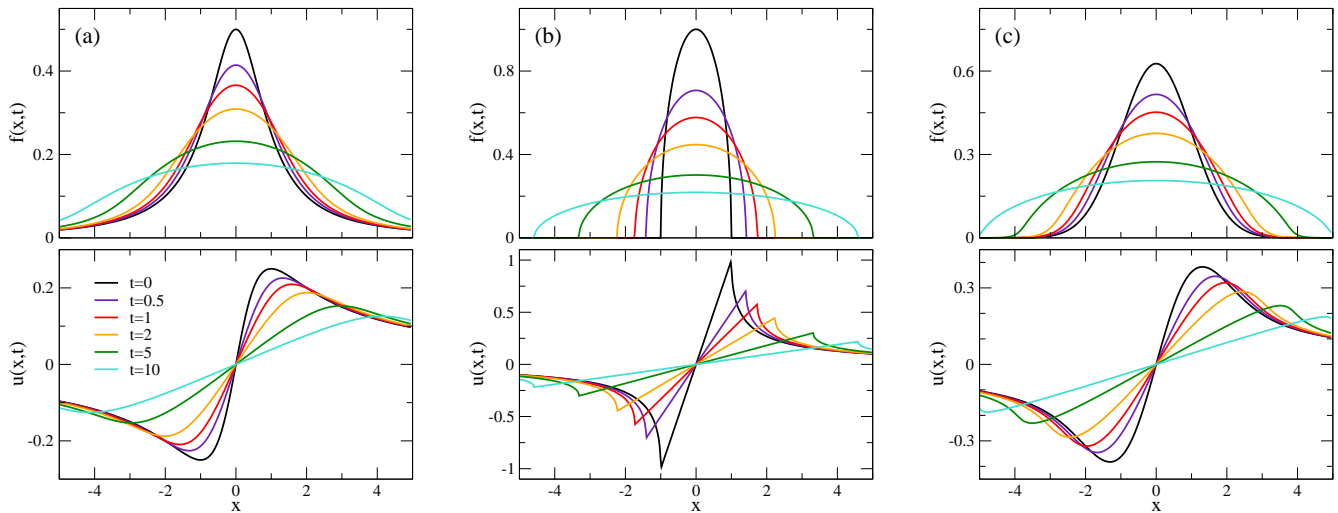

\includegraphics[width=5.4cm]{Fig2a.eps}   \hspace*{0.5cm}
\includegraphics[width=5.4cm]{Fig2b.eps}   \hspace*{0.5cm}
\includegraphics[width=5.4cm]{Fig2c.eps}
\caption{
Spreading of a surfactant pulse: 
Concentration and velocity profiles for a Cauchy, circular and Gaussian pulse, from  left to right. 
The typical width is $a=1$ and the total amount of surfactant is the same in all cases. 
Time is $t= 0, 0.5, 1, 2, 5, 10$ from top to bottom.
Note that the vertical scale changes from one graph to the other.}  
\label{fig:pul}
\end{figure*}

\section{Transient spreading at infinite P\'eclet number}
\label{sec:1d}

The situation  most amenable to analytical treatment is 
when diffusion and evaporation are both absent ($\alpha=\bet=0$  or $\Da=\Pe^{-1}=0$). 
The equation to solve in this case is the inviscid Burgers equation~\cite{book_Burgers-NonlinDifEq,book_Whitham-LinearNonlinearWaves,Bonkile_pramanajp-2018}
\begin{align}
\partial_t \Psi  + \Psi \partial_x \Psi = 0.  
\label{eq:eqtransient}
\end{align}
If  $\Psio(x) \eqdef \Psi(x,0)$  denotes the initial value, 
the solution given by the method of characteristics~\cite{book_aw-MathMethodsPhys} is simply
\begin{align}
\Psi(x,t) = \Psio(\us) \ , \quad \text{with} \quad \us +  t  \Psio(\us) = x. 
\label{eq:defus}
\end{align}
It remains only to solve the equation on $\us$.  
We have identified five relevant cases where this is possible: 
two types of surfactant ``pulses'', two types of surfactant ``holes'' 
and one periodic distribution.

\subsection{Spreading dynamics of surfactant pulses}
\label{pulse}

\paragraph*{Cauchy pulse.}
We first assume that the initial profile is a Cauchy (or Lorentzian) distribution 
with amplitude~$\am$ and width~$a$. The initial values are then
\begin{align}
 \fo(x)  =\frac{\am a^2}{a^2+x^2},  \quad \text{and} \quad  \hifo(x) =\frac{x}{a}\fo(x), 
\end{align}
so that $\Psio(x) =a\am/(x+ i a)$.
Given the simplicity of the expressions,  
Eq.~\eqref{eq:defus}  reduces to a second order equation for $\us$
\begin{align}
 {\us}^2 - (x-ia) \us +  \am a t -i a x  = 0,
\end{align}
which leads to the solution 
\begin{align}
 \Psi(x,t) = \frac{1}{2 t} \left( x + i a - i  \left[ (a-ix)^2+4 a \am t\right]^{1/2}  \right). 
\end{align}
Separating the real and imaginary parts~\footnote{To separate the real and imaginary part, one can use 
the relation $\sqrt{2} \sqrt{x+i y}  =\sqrt{\rho + x} + i \sign(y) \sqrt{\rho -x}$, where $x$ and $y$ are reals and $\rho=\sqrt{x^2+y^2}$.}, 
the concentration and velocity profiles are  obtained explicitly as
\begin{subequations}
\begin{align}
  f(x,t)        &=       \frac{1}{ 2 t} \left(  \Upsilon^{+}(\xi) - a \right),    \label{eq:fpulse}                 \\ 
  u(x,t)        &=       \frac{1}{ 2 t} \left(  x - \Upsilon^{-}(\xi) \right),                           
\end{align}
\end{subequations}
where we define $\xi \equiv   a^2+ 4 a\am t-x^2$  and 
\begin{align}    
 \Upsilon^{\pm}(\xi) &\equiv \sqrt{\left( \sqrt{\xi^2 +  (2 a x)^2}\pm \xi \right) /2} .
\end{align}
To keep compact expressions, we have assumed $x>0$. 
The negative part can be obtained by symmetry 
since $f(x,t)$ and $u(x,t)$ are respectively even and odd  with respect to~$x$. The resulting concentration and velocity profiles are shown in Fig.~\ref{fig:pul}(a).  

\paragraph*{Circular pulse.}
As a second instance, 
we consider an initial surfactant profile that is elliptical or 
``circular'' after rescaling:   
\begin{subequations}
\begin{align}
\fo(x)   &=       \am     \pos{1-x^2/a^2}^{1/2},                      \\ 
\hifo(x) &= \frac{\am}{a} \left( x - \pos{x^2-a^2}^{1/2} \right),                         
\end{align}
\end{subequations}
with  radius~$a$ and amplitude~$\am$. 
The brackets $\pos{.}$ indicate the positive part of the argument. 
Here  we also assume  $x>0$ for the sake of brevity.  
The initial values lead to $\Psio(x) =  \left( \sign(a-x) \sqrt{a^2-x^2} + i x \right)\am/a$.
Equation~\eqref{eq:defus}  is again second order for $\us$ 
and proceeding as above, the concentration and velocity profiles then read
\begin{subequations}
\label{eq:solcirpulse}
\begin{align}
 f (x,t) &=  \frac{a \am}{a(t)}    \pos{1 - \frac{x^2}{a^2(t)}}^{1/2}, \\
 u (x,t) &=  \frac{a \am}{a^2(t) } \left(   x -  \pos{x^2 - a^2(t)}^{1/2} \right),
\end{align}
\end{subequations} 
where $a(t)=\sqrt{a^2+ 2 a \am t}$. 
The circular solution, plotted in Fig.~\ref{fig:pul}(b), is  self-similar. 
Indeed, Eq.~\eqref{eq:solcirpulse} shows that the profile at any time can be obtained by 
replacing the initial radius~$a$ with a time-dependent  radius~$a(t)$,  
while keeping a fixed  amount of surfactant.  
In the context of Marangoni spreading, 
the circular solution was initially found  by Thess~\cite{Thess_ps-1996} with an adhoc method.   
In the mathematical literature~\cite{Biler_crasp-2011}, 
it is known  as a fundamental solution which arises when assuming  self-similarity.    
Note that like other self-similar solutions, such as the Barenblatt-Pattle solution for the porous medium equation~\cite{book_Barenblatt-InterAsymptotics,Pattle_qjmam-1959}, 
the circular pulse solution  has a  finite support.  

\paragraph*{More general pulse shapes.}
The Cauchy and circular pulses are special because they both lead to a second order equation for Eq.~\eqref{eq:defus}.  
Other pulse shapes yield an equation on~$\us$ that in general cannot be solved explicitly~\footnote{
The derivative of a Cauchy pulse leads to a third-order equation that is also solvable.}. 
Still, 
it is always possible to resort to numerical resolution. 
The example of a Gaussian pulse  treated this way  is shown in Fig.~\ref{fig:pul}(c). 
In the  limit $t \rightarrow \infty$, 
all pulses eventually approach the fundamental circular solution~\footnote{
In the sense that the norm of the difference between 
the solution and the circular solution tends toward zero at large time, 
that is 
$||f(x,t) - f_\mathrm{c}(x,t) ||_p \to 0$ for $t \to \infty$, with $||\cdot ||_p$ a $p$-norm.}.
However, 
it is apparent that such loss of memory of the initial profile happens only at long time, 
for $t \simeq 10$ in dimensionless units.  
This indicates  that the shape of the initial pulse has a significant influence on transient Marangoni spreading.  

\paragraph*{Effective diffusion coefficient.}
We now examine whether the spreading process can be described as an effective diffusion process, 
as suggested in several studies~\cite{Suematsu_langmuir-2014,Kitahata_jcp-2018,Xu_csa-2022}. 
For the circular pulse, 
the spatial extension squared $a^2(t)=a^2+ 2 a \am t$ increases linearly with time, a feature typical of diffusion. 
But for the Cauchy pulse, 
the second moment of the concentration profile is infinite 
so that the width of the surfactant distribution is not properly defined.  
One can use instead the position~$\xvmax$  where the velocity is  maximal. 
In the large time limit, one  finds $\xvmax \simeq 2 \sqrt{a \am t}$, 
a behavior again reminiscent of diffusion. 
An effective diffusion coefficient can thus be defined as $\Deff \sim a \am$.
Interestingly, $\Deff$ is 
proportional to the total amount of surfactant in the pulse. 
Thus, the more surfactant in the pulse, the faster the spreading. 
Such a feature, which would not be permitted with plain diffusion 
since the corresponding  equation is linear, 
is a consequence of  nonlinearity.  
Surfactant spreading may thus be described as a diffusive process, 
but only under specific circumstances and with regard to the time evolution of the concentration profile.

\begin{figure*}[t!]
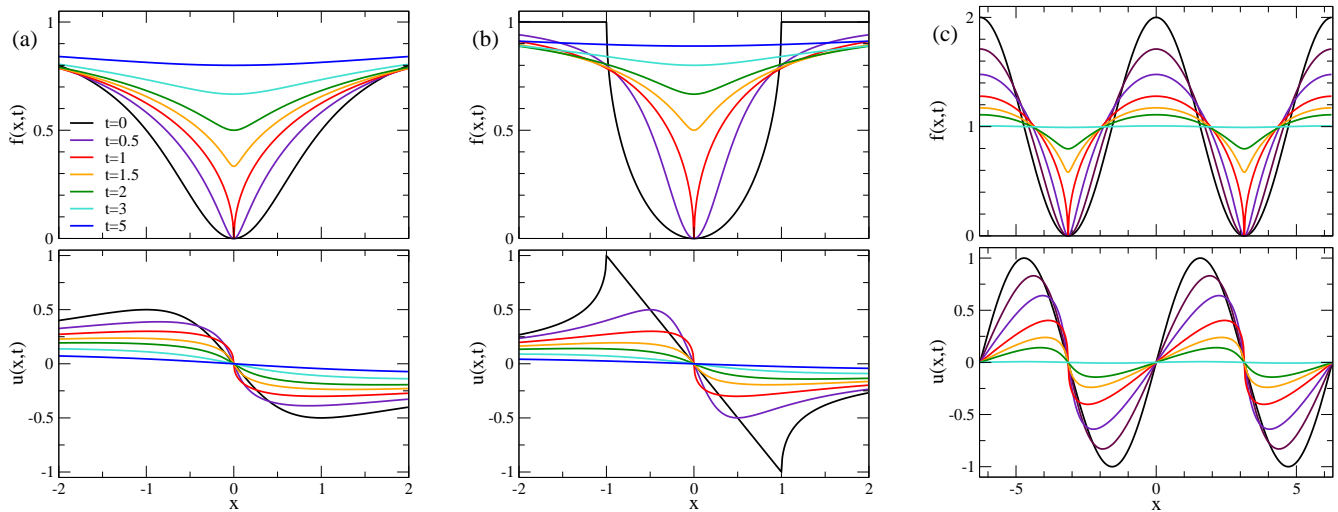

\includegraphics[width=5.4cm]{Fig3a.eps} \hspace*{0.5cm}
\includegraphics[width=5.4cm]{Fig3b.eps} \hspace*{0.5cm}
\includegraphics[width=5.4cm]{Fig3c.eps}
\caption{
(a-b) 
Closing of a surfactant hole:
Concentration and velocity profiles for Cauchy and circular holes.
Here $\cinf=a=\am=1$. 
Time is $t=0, 0.5, 1, 1.5, 2, 3, 5$ from  bottom to top. 
A singularity occurs at $\tsing=1$ and $\xsing=0$.
(c)
Relaxation of a periodic distribution of surfactant: 
Concentration and velocity profiles for a sinusoidal initial distribution, 
with $\cinf=a=\am=1$. 
Time is $t=0, 0.2, 0.5, 1, 1.5, 2, 5$.
A singularity occurs at $\tsing=1$ and $\xsing=\pm \pi$.}  
\label{fig:holper}
\end{figure*}

\subsection{Closure dynamics of surfactant holes}
\label{holes}

We now discuss the reverse situation where 
surfactant is already present at the interface 
but with a concentration near the origin that is lower than the value $\cinf$ far away  from it. 
Such surfactant ``dimple'' or for brevity ``hole'' in the following 
is expected to close as a result of Marangoni flow.  
As for the spreading case, 
we derive two exact solutions for this process: the Cauchy hole and the circular hole. 

\paragraph*{Cauchy hole.} 
Consider the initial surfactant concentration given by
\begin{align}
\fo(x) = \cinf  - \frac{a^2 \am}{a^2+x^2}, 
\end{align}
with $0 \leqslant\am \leqslant \cinf$. 
Retracing the steps detailed above, 
the $\Psi$ function, concentration and velocity can all be written explicitly in a straightforward manner. 
For the sake of clarity, 
the mathematical expressions are reported in App.~\ref{sec:appformula}. 
An interesting feature occurs when the initial concentration vanishes at the origin, 
\ie{} when $\am =\cinf $.
In this case, the concentration profile exhibits a  singularity at a finite time $\tsing=a/\cinf$. 
Indeed, a small-$x$ expansion  gives at lowest order
\begin{align}
  x \rightarrow 0 , \quad   f(x,\tsing)  = - \frac{u(x,\tsing)}{\sign(x)} = \frac{\cinf}{\sqrt{2 a}} \sqrt{|x|}.    \label{eq:holesingularity} 
\end{align}
This indicates that the concentration profile has a cusp near the origin, as illustrated in Fig.~\ref{fig:holper}(a). Regarding the velocity profile, it has an infinite slope at the origin when $t= \tsing$.  
For larger time $t > \tsing$, the profile becomes regular again.  
Note that this singularity has already been identified in previous investigations~\cite{Crowdy_siamjam-2021}, 
but focusing on the dynamics of  poles in the complex plane.   
For $\cinf = \am$, the concentration at the origin  has a remarkably simple expression 
\begin{subequations}
\begin{align}
 t \leqslant \tsing, \quad f(0,t) &= 0,  \\
 t \geqslant \tsing, \quad f(0,t) &= \cinf \left( 1 - \frac{\tsing}{t}\right). 
\end{align}
\end{subequations}
The concentration remains zero until the cusp forms at time~$\tsing$,  
after which it relaxes as $t^{-1}$.

\paragraph*{Circular hole.} 
The case of an elliptic initial dimple can also be treated analytically. The initial density reads
\begin{align}
\fo(x) = \cinf  - \frac{\am}{a} \pos{a^2-x^2}^{1/2}, 
\end{align}
with the condition $0 \leqslant\am \leqslant \cinf$. 
The exact solution is provided in App.~\ref{sec:appformula}.  
The concentration and velocity profiles are displayed in Fig.~\ref{fig:holper}(b). 
When $\am=\cinf$, the concentration exhibits a finite-time singularity as well, 
with features similar to those discussed above.   
In particular, Eq.~\eqref{eq:holesingularity} still holds, with the factor of~2 removed. 
Regarding the concentration at the origin, one now gets
\begin{subequations}
\begin{align}
 t \leqslant \tsing, \quad f(0,t) &= 0,  \\
 t \geqslant \tsing, \quad f(0,t) &= \cinf \left( 1 - \frac{\tsing}{2t-\tsing}\right).  
\end{align}
\end{subequations}
We therefore find that the concentration relaxes asymptotically with the same $t^{-1}$ law.

\subsection{Periodic distribution of surfactant}

We now turn to a periodic initial distribution of surfactant 
with sinusoidal variations  of period~$ 2 \pi a$ and amplitude~$\am$ around the mean value~$\cinf$~\footnote{For the concentration to remain positive, one should have $0 \leqslant\am \leqslant \cinf$.}.   
From $\fo(x) =\cinf + \am \cos(x/a)$ and  $i \Psio(x) =\cinf + \am \exp(i x/a)$,  
the solution  can be found  as 
\begin{align}
i \Psi(x,t) = \cinf  + \frac{a}{t} W\left( \frac{t \am}{a} \exp\left[ (i x - \cinf t)/ a \right] \right), 
\label{eq:solper}
\end{align}
where $W$ is the principal value of the Lambert function
that satisfies $W(x) \exp\left[W(x)\right] = x$~\cite{Corless_acm-1996}.  
The corresponding profiles are plotted  in Fig.~\ref{fig:holper}(c). 
Once again, if the initial concentration includes a point of vanishing concentration (\ie~$\am=\cinf$), 
the profile exhibits a singularity  at  $(\tsing,\xsing)=(1,\pm \pi a)$ 
with the same features as discussed previously.  
As a side remark, we note that the solution of Eq.~\eqref{eq:solper} derived for an infinite system 
also applies to a finite domain~$x \in [-\pi a, \pi a]$ with no-flux boundary conditions. 

The asymptotic behavior can also be extracted from Eq.~\eqref{eq:solper}.
In the large-time limit, the relaxation is found to be exponential  
\begin{align}
 t \rightarrow \infty, \quad f(x,t) - \cinf  = \am \cos(x/a) e^{-t/\taur}.   
\end{align}
The characteristic time $\taur=a/\cinf$  is independent of the modulation amplitude~$\am$ 
but  decreases with the mean concentration~$\cinf$. 
In other words, 
a surfactant-laden interface can erase initial inhomogeneities  more rapidly when richer in surfactant.  
At this point, it is instructive to compare with a purely diffusive process. For diffusion, 
the relaxation of a sinusoidal profile 
is exponential at all time with a characteristic time~$\tau_D=a^2/D$. 
Here, Marangoni spreading exhibits a similar behavior: setting $\taur=a^2/\Deff$, one can define an effective diffusion coefficient $\Deff \sim a \cinf$ 
that is again proportional to the total amount of surfactant involved, 
as already discussed in Sec.~\ref{pulse}.  
The analogy with diffusion is, however, incomplete.  
As clearly seen in Fig.~\ref{fig:holper}(c), 
 the concentration profile in surfactant-poor and surfactant-rich regions 
evolves in a very asymmetric manner. 
Such a feature, that arises from the nonlinearity of Eq.~\eqref{eq:eqtransient},  
would be proscribed in a diffusion process.

\section{Spreading dynamics at finite Damk\"{o}hler or P\'eclet numbers}
\label{sec:1d_diff}

When either evaporation or diffusion are relevant, 
additional terms have to be accounted for 
and the inviscid Burgers Eq.~\eqref{eq:eqtransient} has to be replaced 
by its more general version Eq.~\eqref{eq:burgers}. 
We discuss in this section the new solutions that arise when considering finite Damk\"{o}hler or P\'eclet numbers.   

\subsection{Effect of evaporation}

For finite Damk\"{o}hler number, \ie{} when evaporation is considered,  
the equation to solve is 
$\partial_t \Psi  +  \Psi \partial_x \Psi = -\alpha \Psi$.
One can easily check that the solution $\Psi$ with evaporation ($\alpha > 0$) is directly related to the solution $\Psidag$  without evaporation ($\alpha = 0$) by
\begin{align}
 \Psi(x,t) = e^{-\alpha t} \Psidag(x,\taueff), \quad \taueff = \frac{1 -e^{-\alpha t} }{\alpha}. \label{eq:solevaporation}
\end{align}
Note in particular that
combining Eq.~\eqref{eq:solevaporation} and  the circular pulse solution Eq.~\eqref{eq:solcirpulse} 
yields back the solution obtained previously in Ref.~\cite{Bickel_sm-2019}. 

Evaporation has two effects. 
First, 
both  the concentration and the velocity  include an exponential decay with rate~$\alpha$. 
Second, 
the effective time~$\taueff$, which is equal to~$t$ at short times~$t\ll \taueff$, 
subsequently reaches a plateau at longer times~$t\gg \taueff$. 
The plateau value $\tau = \alpha^{-1}$ corresponds to the characteristic time associated with evaporation. 
This implies that  the long-time behavior of the relaxation is always exponential, whatever the initial distribution.  
Asymptotically, evaporation is thus the dominant transport mechanism. 

\subsection{Effect of surface diffusion}

\subsubsection{Fundamental solution}
We come back to the situation where there is no evaporation and focus on the effect of surfactant diffusion 
along the interface.  
We assume that  the interface is initially clean, 
with $f(x,t=0) \to 0$ for $x \to \infty$. 
The equation to solve is now the  viscous Burgers equation
\begin{align}
\partial_t \Psi  + \Psi \, \partial_x \Psi =  \eps \, \partial^2_{xx} \Psi,
\end{align} 
with $\eps$ the inverse P\'eclet number which is proportional to the surface diffusion coefficient~$\Ds$.
This nonlinear equation can actually be converted to a linear partial differential equation
thanks to the Hopf transformation~\cite{Hopf_cpam-1950}.
Denoting as $\Psio(x)$ the initial value, 
the general solution can  be expressed as 
\begin{align}
 \Psi(x,t) &= - 2 \eps  \, \partial_x \log I(x),  \label{eq:Phivisburgers}                            
\end{align}
where we define
\begin{align}
      I(x) &\equiv    \intii   \exp \left( -g(u)/\eps \right) \dif{u}, \\                   
      g(u) &\equiv   \frac{(u-x)^2}{4t} + \frac{1}{2} \int_0^u \Psio(u') \,\dif{u'}.   
      \end{align}
This general solution is now specified for a Cauchy pulse  
because among the initial profiles we consider, 
this case is the only one we found to be fully tractable analytically.   
Considering the initial value $\Psio(x) = a\am/(x+i a)$, one  gets 
\begin{align}
 I(x) = \intii \left(1 - i u/a \right)^{-\frac{a\am}{2\eps}} \exp\left[ -\frac{(u-x)^2}{4 \eps t} \right]\, \dif{u}. 
 \label{eq:I(x)}
\end{align}
While the result of the latter integral is available for any width~$a$ and amplitude~$\am$~\footnote{
A Hubbard-Stratonovitch transform is first introduced for the exponential term. 
After switching the integration order, the double integral can also be obtained.}, 
we focus for simplicity on the limit 
$a \rightarrow 0$ while keeping $a\am=1/\pi$, 
so that the total amount of surfactant is unity. 
In this limit, the initial Cauchy pulse approaches a Dirac distribution. 
The explicit expression for $I(x)$ is then 
\begin{align}
 I(x) =& \Gamma\left( \nuexp \right)  \,_1F_1 \left( \nuexp, \frac{1}{2}, - \frac{x^2}{4 \eps t} \right)  + \nonumber \\ 
       & i \frac{2  x}{\sqrt{4 \eps t}} \,  \Gamma\left( \frac{1}{2} +\nuexp \right)  \,_1F_1 \left( \frac{1}{2}+\nuexp, \frac{3}{2}, - \frac{x^2}{4 \eps t} \right), 
\end{align}
where $\Gamma$~denotes here the gamma function, 
$_1F_1$~is the confluent hypergeometric function~\cite{book_GradshteynRyzhik-TableInt} and 
$\nuexp \eqdef 1/4\pi\eps$. 
Using Eq.~\eqref{eq:Phivisburgers}, 
fully explicit expressions can be written for 
the velocity and concentration profiles. 
The resulting  formulas then constitute the fundamental solution 
for Marangoni spreading with surface diffusion. 

\begin{figure}[t!]
\includegraphics[width=8.5cm]{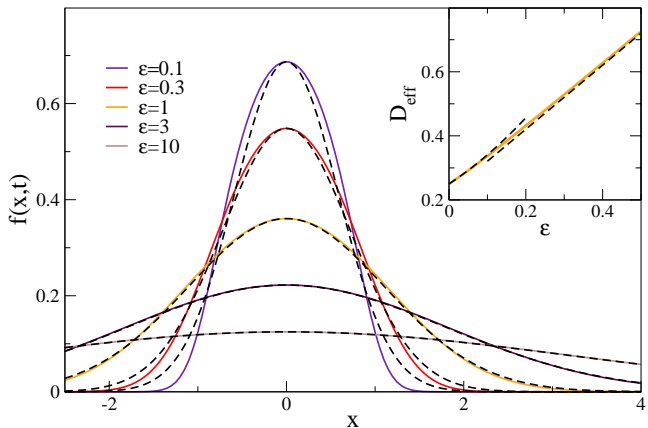} 
\caption{
Influence of surface diffusion on the transient spreading of a Dirac pulse. 
Here, time is fixed to $t=0.5$ and 
the concentration profile is shown for $\eps=0.1, 0.3, 1, 3, 10$ from top to bottom.
The dashed lines correspond to Gaussian distributions with variance $2 t \Deff(\eps)$.
Inset:   
Effective diffusion coefficient $\Deff(\eps)$ as given by Eq.~\eqref{eq:Deff}. 
Approximations from Eqs.~\eqref{eq:Deffsmalleps} and ~\eqref{eq:Defflargeeps} 
are also shown  with dashed lines.
}
\label{fig:pulsediffusion}
\end{figure}

\subsubsection{Effective diffusion coefficient}
The primary effect of diffusion is to smooth out the distribution. 
This  is illustrated in Fig.~\ref{fig:pulsediffusion} at fixed time:  
the  stronger the diffusion coefficient, the wider the surfactant distribution. 
Because the spreading of a Cauchy pulse is already diffusive-like when $\eps=0$, 
one  expects that
the whole process --- now including genuine surface diffusion ---  
can be described with an effective diffusion coefficient. 
To do so, 
we focus on the concentration at the origin, which is  given by the remarkably simple expression
\begin{align}
f(x=0,t) = \frac{1}{\sqrt{4 \pi \Deff t}}, \label{eq:corigin}
\end{align} 
with an effective coefficient~$\Deff$ defined as
\begin{align}
 \Deff (\eps)= \pi \eps \, \left[ \frac{ \Gamma \left(1            + \nuexp \right)} 
                                  {\Gamma \left( \frac{1}{2} + \nuexp \right)}  \right]^2, 
                                  \quad \nuexp \eqdef 1/4\pi\eps. 
                                  \label{eq:Deff}
\end{align}
Equation~\eqref{eq:corigin} is  valid at all time and 
matches rigorously what is expected for a purely diffusive process. 
A similar dependence also holds asymptotically at long time for any finite position~$x$.  
The effective diffusion coefficient~$\Deff (\eps)$ is plotted as a function of~$\eps$ in the inset of Fig.~\ref{fig:pulsediffusion}. 
We can identify the limiting behaviors
\begin{align}
 \eps \rightarrow 0,      & \quad \Deff(\eps) = \frac{1}{4} \left( 1 + \pi \eps +  \frac{\pi^2}{2} \eps^2 +...\right),  \label{eq:Deffsmalleps}   \\
 \eps \rightarrow \infty, & \quad \Deff(\eps) = \eps + C_1,                                                             \label{eq:Defflargeeps}
\end{align}
where $C_1 \approx 0.22$ is a numerical constant~\footnote{
$C_1 = -(\gamma_\mathrm{E} +\psi^{(0)}(1/2))/2\pi$, 
with $\gamma_\mathrm{E}$ Euler's constant and $\psi$ a polygamma function.}. 
One thus recovers the expected values in two limits.  
For very weak diffusion $\eps \ll 1$, 
the value of the effective coefficient approaches $\Deff=1/4$, 
in agreement with Eq.~\eqref{eq:fpulse} for the zero-diffusion case~\footnote{Remember 
that we set $a \am=1/\pi$ and take the limit $ a \to 0$.}. 
On the other hand,  
one obtains $\Deff \simeq \eps$ when intrinsic diffusion dominates.  
Finally,  
since a Gaussian behavior is expected for pure diffusion, 
the concentration profiles can be compared  to Gaussian distributions 
with variance $2 t \Deff(\eps)$, 
as done in Fig.~\ref{fig:pulsediffusion}.   
Though small discrepancies are visible in the tails of the distributions, 
the agreement is fairly good.  
The fundamental solution of Marangoni spreading  
may thus be reasonably approximated as a diffusive process 
with effective coefficient~$\Deff(\eps)$ given by Eq.~\eqref{eq:Deff}.

We have focused so far on surfactant spreading in the transient regime. 
Yet another situation of interest is when the surfactant is released 
continuously~\cite{Roche_prl-2014,LeRoux_pre-2016,Mandre_jfm-2017,Kitahata_jcp-2018,Arangalage_sm-2018,Benouaguef_jfm-2021}. 
We describe in App.~\ref{sec:appsteady} the solutions that 
are available for this steady state source in the one-dimensional case.

\section{Axially symmetric spreading in two dimensions}
\label{sec:2d}

\subsection{Closure relation and Riesz transform} 

In this section, 
we briefly discuss transient surfactant spreading in higher dimensionality.  
Evaporation and diffusion are discarded everywhere.  
Space dimensionality is noted $D$  and the dimension of the interface is $d=D-1$.
Keeping the assumptions  made in Sec.~\ref{sec:model},  
the  closure relation of Eq.~\eqref{eq:closure1D} can be generalized in any space dimension as~\cite{Thess_jfm-1997}
\begin{align}
 \vus (\vx)       &= -\bm{\nabla} (-\Delta)^{-1/2}       f(\vx),   \label{eq:uxfromfx} 
\end{align}
or, equivalently, in Fourier representation
\begin{align}
 \fou{\vu}_\mathrm{s} (\vq) &= - \frac{i\vq}{q}          \fou{f}(\vq).   \label{eq:uqfromfq}
\end{align}
Here, $\vus$ denotes the in-plane velocity at the interface, 
$\vx$~is the position along the interface   
and $\fou{f} \eqdef \fouopbra{f}$ is the Fourier transform of a 
function~$f$~\footnote{Equation~(\ref{eq:uqfromfq}) depends on the convention chosen for the definition of Fourier transform.}.   
For the dimension $d=1$ discussed until now,   
the velocity is given by the Hilbert transform of the concentration. 
For the dimension $d=2$ considered in this section,  
the velocity can be expressed as the Riesz transform of the concentration,  $\vu_s = \rizopbra{f}$,  where we define
\begin{align}
\rizopbra{f}(\vx) \equiv \frac{1}{2\pi} \PV \int  \frac{\vx - \vx'}{|\vx - \vx'|^3} f(\vx') \dif{\vx'}.   
\end{align}
Consequently, the closure relation is again non-local. 

Whereas an extensive literature can be found on Hilbert transforms, 
the use of Riesz transforms appears to be less common. 
Moreover, 
it does not seem possible to recast the problem as a Burgers equation. 
From now on, we  restrict the discussion to radially symmetric pulses, 
such that the concentration~$f$ depends only on  the distance~$r$ to the origin. 
Denoting as~$\mathbf{e}_r$ the radial unit vector,  
the velocity is also radial with component $u_r=\rizopbra{f} \cdot \mathbf{e}_r \eqdef \riz{f}$.   
The equation governing the evolution of surfactant distribution~$f(r,t)$ is then  
\begin{align}
 \partial_t f +  \frac{1}{r} \partial_r \left( r f  \riz{f} \right) =0.  
  \label{eq:eq2Dinsolubleradial} 
\end{align}
Even within these assumptions, 
computing the Riesz transform is in general not straightforward. 
Several expressions that are useful for this purpose are given in App.~\ref{sec:appriesz}. 

\subsection{Transient spreading}

By analogy with the one-dimensional case,  
we consider the spreading of a circular pulse
\begin{align}
 f(r,t) = \frac{a^2 \am}{a^2(t)} \left[ 1 - \left( \frac{r}{a(t)}\right)^2 \right]^{1/2}_+, 
 \label{eq:f2Dcirpulse}
\end{align}
where  $a$ and $\am$ are respectively the initial radius and amplitude, 
and the time-dependent radius~$a(t)$ is to be determined.
Computing the Riesz transform, we get
\begin{align}
 \riz{f}(r,t) & = \frac{\pi a^2 \am}{4 a^3(t)}  r,    \label{eq:u2Dcirpulse}      \\
 \riz{f}(r,t) & = \frac{ a^2 \am}{2 a^3(t) }  \left[  - \frac{a(t)}{r} \sqrt{ r^2-a^2(t)}   + r \arcsin  \left( \frac{a(t)}{r}\right) \right].   \nonumber
\end{align}
for $ r \leqslant a(t)$ and $r \geqslant a(t)$ respectively. 
Equation~\eqref{eq:eq2Dinsolubleradial} is then satisfied provided $a'(t) a^2(t) = - \pi a^2 \am/4$, 
which finally  leads to
\begin{align}
 a(t) = \left( a^3 + \frac{3 \pi a^2 \am}{4}  t \right)^{1/3}. 
\end{align}
We recover the self-similar solution that was 
identified in the mathematical literature~\cite{Biler_crasp-2011} 
but using a completely different method. 
The transient spreading in the two-dimensional geometry is therefore subdiffusive, 
since the spatial extent  increases as $a(t) \sim t^{1/3}$.  

\begin{figure}[t!]
\includegraphics[width=8cm]{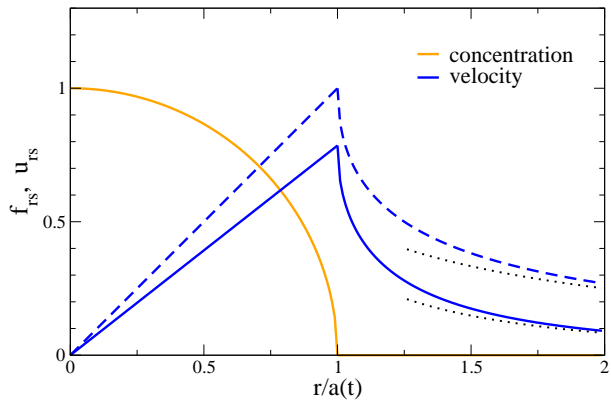} 

\caption{Concentration and velocity profiles for the circular solution to transient spreading. 
Shown are  
the rescaled concentration $f_\mathrm{rs} \eqdef \lambda(t) f(r,t)$ 
and the rescaled velocity  $u_\mathrm{rs} \eqdef \lambda(t) u(r,t)$, 
with a scaling factor $\lambda(t) = (a(t)/a)^d /\am$, 
$d$ the dimension, $a(t)$ the radius at time~$t$ and $a$ the initial radius.   
The case $d=2$  (solid lines) and $d=1$ (dashed line)  correspond respectively to 
Eqs.~\eqref{eq:f2Dcirpulse}-\eqref{eq:u2Dcirpulse} and Eq.~\eqref{eq:solcirpulse}. 
The dotted line is the large distance approximation of velocity from Eqs.~\eqref{eq:u1Dcirpulselargex} and \eqref{eq:u2Dcirpulselarger} 
taken at lowest order. }
\label{fig:pul2d}
\end{figure}

It is interesting to compare the characteristics of the circular pulse for $d=1$ and $d=2$. 
As visible in Fig.~\ref{fig:pul2d}, 
both solutions have a finite support within which the velocity is linear 
and also feature an angular point with infinite slope at the boundary $r=a(t)$. 
At distances much larger than $a(t)$, the velocity behaves as  
\begin{align}
d=1, \quad \frac{u(x)}{a   \am} &= \frac{1}{2 x} + \frac{a^2 + 2 a \am t}{8 x^3} +...,  \label{eq:u1Dcirpulselargex}   \\ 
d=2, \quad \frac{u(r)}{a^2 \am} &= \frac{1}{3 r^2} + \frac{ a^2(t)}{10 r^4} + ...      \label{eq:u2Dcirpulselarger} 
\end{align}
The velocity is thus proportional to the ``volume'' of the  pulse $\sim a^d \am$ and 
decays as a power law~$u(r)\sim r^{-d}$. 

\subsection{Steady spreading from a source}

As the last solvable case,  
let us consider a punctual source releasing a steady flux of surfactant~$Q$ 
that spreads in two dimensions, without evaporation. 
Assuming steady state, 
 Eq.~\eqref{eq:eq2Dinsolubleradial} or surfactant conservation gives
\begin{align}
2 \pi r \, f(r) \, \riz{f}(r) = Q.             \label{eq:eq2Dsourcenoevap}
\end{align}
Using the Riesz transform expressions of App.~\ref{sec:appriesz}, 
one can show that a solution is
\begin{subequations}
\begin{align}
f(r) &= \sqrt{\frac{2}{\pi}} \frac{\Gamma \left(\frac{5}{4}\right) }{\Gamma \left(\frac{3}{4}\right) } \,  \sqrt{\frac{Q}{r}},  \\
u(r) &= \sqrt{\frac{2}{\pi}} \frac{\Gamma \left(\frac{3}{4}\right) }{\Gamma \left(\frac{1}{4}\right) } \,  \sqrt{\frac{Q}{r}}.   
\end{align}
\end{subequations}
The concentration $f$ and the velocity $u=\riz{f}$ display the same power-law dependence,  
a behavior clearly distinct from diffusive behavior where no steady state exists when $d=2$.  
The $r^{-1/2}$ dependence might be understood from a simple argument. 
Since there is no characteristic length scale other than the distance to the source, 
one can expect a typical viscous stress $u(r)/r$. 
On the other hand, the typical Marangoni stress is $f'(r)$. 
Equating viscous and Marangoni stresses and using Eq.~\eqref{eq:eq2Dsourcenoevap}, 
one finds $f'(r) f(r) \sim r^{-2}$ and $f(r) \sim u(r) \sim r^{-1/2}$.

For a steady source of insoluble surfactant~\cite{Mandre_jfm-2017,Bandi_prl-2017}, 
Mandre found that the velocity field decays as $r^{-\nu}$ with~$\nu=3/5$,  
whereas we find an exponent~$\nu=1/2$. 
The former applies when there is a boundary layer for the flow, the latter holds for Stokes flow. 
An other point of comparison is the 
exact solution from Bratukhin and Maurin for a steady punctual source of heat or soluble surfactant in three  dimensions~\cite{Bratukhin_jamm-1967, Shtern_jfm-1993},   
in which the far-field velocity decays as~$r^{-1}$. 
Spreading of an insoluble surfactant in the Stokes regime 
has thus a distinct hydrodynamic signature with the slowest velocity decay.

\section{Discussion } 
\label{sec:dis}

To summarize, we developed a unified analytical approach  to describe 
the transient and steady Marangoni spreading of an insoluble surfactant  
in the Stokes flow limit~\footnote{Whether in a specific experiment the surfactant may actually be considered as insoluble  depends on a number of factors, including the characteristic time for desorption from the interface,  the time over which spreading is observed and the initial concentration of 
surfactant~[44].}.  
In the one-dimensional case, 
the mapping between an intricate set of transport equations and a complex Burgers equation  
is derived in a straightforward manner using the properties of Hilbert transforms. 
The solutions previously uncovered in the various fields of the literature 
-- from applied mathematics to physical chemistry through fluid mechanics -- 
are gathered within a single framework. 
Importantly, 
our approach allows to identify all cases where a fully explicit solution is possible.   
By investigating a number of them, 
we show that the nonlinearity of the Burgers equation may lead to a variety of behaviors in spreading. 
Finally, 
for an initial surfactant distribution with arbitrary shape,   
the solution will not be generally available in analytical form 
but may be obtained numerically by solving a simple equation. 
With a set of exact solutions in hands,   
we can now discuss more thoroughly the physical insights they provide.  
To do so, we consider several points in turn, 
focusing on the one-dimensional case if not mentioned otherwise.

\paragraph*{Time scale for Marangoni spreading.} 
If we switch back to dimensional variables, 
the time scale involved with transient surfactant spreading reads  $\tau=2\eta L/\kap \Gamo$, 
with $L$ the characteristic length scale of the perturbation. 
This time scale is thus inversely proportional to the surfactant concentration. 
Taking $\Gamo \simeq 10^3$~molecules/\textmu m$^{2}$, which corresponds to a low surface fraction $\varphi \simeq 10^{-3}$,  
$\kappa=k_\mathrm{B} T \simeq 4\times 10^{-21}\,$J at room temperature  
and the viscosity of water~$\eta \simeq 10^{-3}\,$Pa$\,$s, 
one gets a macroscopic time $\tau \simeq 1\,$s for $L=1\,$mm. 
Our results are thus relevant with regards to experimental time scales.

\paragraph*{(Dis)similarities between pulses and holes.} 
To compare the complementary situations of pulses and holes, 
we examine the long-time behavior of the concentration~$f(x,t)$, 
which is normalized by the ``volume''~$a \am$ of the initial perturbation 
so that whatever the amount of surfactant, the spatial integral is unity.
For any finite position~$x$,  
the concentration evolves asymptotically as
\begin{align}
t \rightarrow \infty, \quad \frac{f(x,t)}{a \am}  \simeq   \frac{1}{\sqrt{\indcase a \am t}}, 
\label{eq:longtimebehpulse}
\end{align}
for a pulse whereas for a hole
\begin{align}
t \rightarrow \infty, \quad  \frac{\cinf  - f(x,t)}{a \am} \simeq  \frac{1}{\indcase \cinf t}, 
\label{eq:longtimebehhole}
\end{align}
with $\chi$  a numerical constant~\footnote{$\indcase=1$ and $2$ for Cauchy and circular cases respectively.}. 
Two differences can be emphasized.  
{\it (i)}~The time relaxation of pulses is essentially diffusive-like, 
with an effective diffusion coefficient $\Deff \sim a \am$ 
proportional to the total amount of surfactant in the pulse. 
In contrast, hole closing cannot be described as a diffusive process. 
{\it (ii)}~The dynamics of hole closing is much faster than pulse spreading.
This difference arises from the non-linear nature of 
the transport equation. Indeed,  with a linear equation, holes and pulses would evolve in the same way. 
This clearly indicates that any analogy with effective diffusion has to be handled carefully.

\paragraph*{Surfactant spreading on a contaminated surface.} 
Our results also indicate that Marangoni spreading is very sensitive to the initial state of the surface: 
There is an essential difference in the behavior expected  
with a perfectly clean surface 
and  a surface already covered with endogenous surfactants~\cite{Grotberg_japphysio-1995,Sauleda_langmuir-2021}. 
Indeed, the solution for the holes still holds when $\am<0$, 
that is when a surfactant pulse is released 
on a surface with an initial homogeneous concentration~$\cinf$. 
In the long-time limit, 
the spreading dynamics  thus depends on whether the interface is clean or contaminated with surface-active molecules.     
Such a change in behavior 
can be traced back to the  additional term $\cinf \, \partial_x \hi{f}$ appearing in Eq.~\eqref{eq:thesseq}, 
which describes the transport of endogenous surfactants by the Marangoni flow due to the added surfactant pulse. 
Because the distribution  of endogenous surfactants may become inhomogeneous,  
a  contaminated interface with concentration~$\cinf$ 
is not equivalent to  a clean interface with surface tension~$\gamo - \kappa \cinf$.  
In particular, the asymptotic dynamics of Eq.~\eqref{eq:longtimebehpulse} for a clean interface 
cannot be recovered by taking the limit~$\cinf \rightarrow 0$  in~Eq.~\eqref{eq:longtimebehhole}. 
This point is especially relevant for experiments with aqueous solvents. 
Indeed, it may provide a clear signature of interfacial contamination that could be tested experimentally.

\paragraph*{Mode decomposition.} 
Whether the decay of an initial surfactant pulse 
can be decomposed in sinusoidal modes is a natural question.
This issue has been addressed recently in a linearized version of the transport equations~\cite{Mcnair_jfm-2022}. 
Here, we solved the full nonlinear equations.
We note in particular that, 
when the amplitude of the perturbation is small with respect to mean level ($\am \ll \cinf$),  
the surfactant profile indeed remains sinusoidal while decaying exponentially. 
This is expected since this assumption makes the problem linear~\cite{Shardt_jfm-2016}.
On the other hand, nonlinear effects become relevant as soon as  
$\am/\cinf$ is not well below unity  
and the mode decomposition does not apply anymore. 

\paragraph*{Effective diffusion and space dimension.}  
In the one-dimensional geometry, 
we have identified  several cases -- including surfactant pulse  with and without surface diffusion -- 
where the Marangoni spreading can be mapped to a diffusion process, though with specific features. 
For a two-dimensional interface, 
the algebra is much more involved and only two analytical solutions are so far available: 
the spreading dynamics for a circular pulse and the steady state source. 
For the former,  the spreading in the long-time limit is subdiffusive with exponent~$1/3$.  
For the latter, the velocity decays as a power law, at odds with a diffusive behavior. 
Taken together, these findings  point to the impossibility  
of defining an effective diffusion coefficient in the two-dimensional case. 
Since this situation is the most relevant experimentally, this calls for caution. 
Even though the idea of effective diffusion is sometimes invoked,  
it appears difficult, in general, to map a transport process dominated by Marangoni convection 
onto a simple diffusive process.

To conclude, 
we briefly discuss how our findings may be tested experimentally.  
Predictions for Marangoni spreading have long been confronted to experiments~\cite{Matar_sm-2009}, 
using for instance fluorescent surfactants~\cite{Fallest_njp-2010,Usma_langmuir-2022}. 
To reproduce the situations considered in this work, 
two peculiar features are needed.  
First, 
the set-up must allow to tailor the initial profile of surfactant to a prescribed shape. 
Second, 
the one-dimensional geometry is preferable because it is best understood and amenable to complete predictions. 
The controlled deposition of surfactant at the interface might be a possible option. 
However, 
the use of photoswitchable surfactant molecules~\cite{Eastoe_sm-2005,Liu_jcis-2009,Chevallier_sm-2013,Kavokine_acie-2016}
could be an ideal method to induce the pulse, hole and periodic patterns that we have considered theoretically. 
Indeed, 
because the activated surfactants appear right at the interface, 
the perturbations  are  minimal.  
Besides, the one-dimensional geometry is easily imposed 
and the concentration profile can be prescribed through the light intensity. 
We thus hope that the exact solutions found in this work can be put to experimental test, 
so as to provide, for the one-dimensional case at least,  
a complete picture for the viscous Marangoni spreading of an insoluble surfactant.

\appendix 

\section{Steady release of surfactant}
\label{sec:appsteady}

\begin{figure*}[t!]
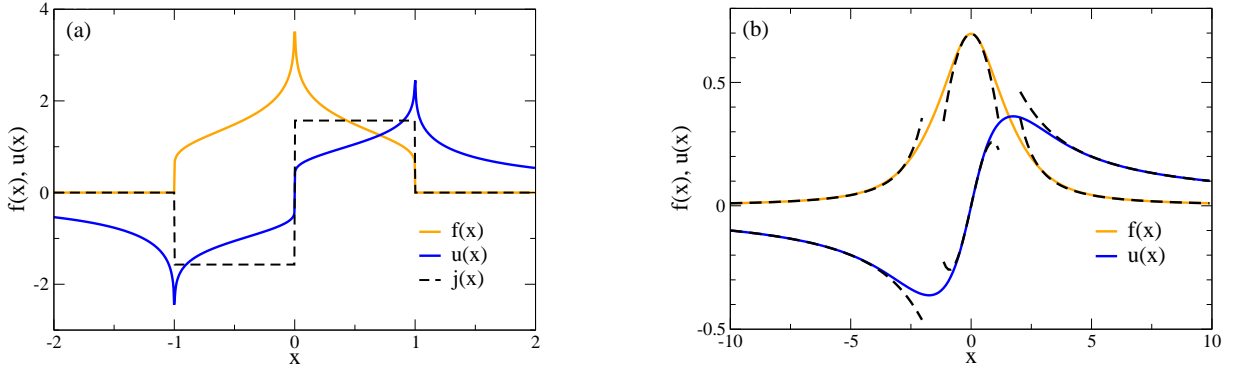

\includegraphics[height=5cm]{Fig6a.eps} \hspace*{1.5cm}
\includegraphics[height=5cm]{Fig6b.eps} 
\caption{
Steady state around a  source of surfactant. 
(a)
Without evaporation. 
The punctual source at the origin is surrounded by two punctual sinks located  at $x=l=\pm 1$. 
The dashed line is the local flux of surfactant $j(x)=u(x) f(x)$. 
(b) With evaporation.
The source at the origin is of Cauchy type, with $a=1$ and $\alpha=1$. 
The dashed lines show the approximations at small and large~$x$.
}
\label{fig:steadystate}
\end{figure*}

We seek a steady state in the plane flow geometry 
when the surfactant is continuously released at the surface 
with a distributed source profile~$q(x)$. 
Introducing the complex flux $\qflux=\hi{q} - i q$ and neglecting diffusion, 
the equation to solve is now
\begin{align}
\Psi \partial_x \Psi = - \alpha \Psi  + \qflux(x).    \label{eq:steadysource}
\end{align}
As before, the source profile  more susceptible to analytical treatment 
is a Cauchy profile  and we set $\qflux = 1/(x + i a)$. 

\paragraph*{No evaporation.}
For a single Cauchy source, 
the solution to Eq.~\eqref{eq:steadysource} diverges for $x \rightarrow \infty$, 
suggesting that there is no steady state.  
Let us assume then that some local surfactant sink exists, 
where surfactant molecules  disappear. 
This is physically realizable if a photoswitchable surfactant can be instantaneously
disabled with a certain light. 
We can thus consider a source surrounded by two sinks at position~$\pm l$, all of Cauchy type. 
The resulting steady state is described by
\begin{align}
\Psi(x) &= - i \sqrt{\Log (l^2+(a - i x)^2) - 2 \Log(a-i x) }.   
\end{align}
In the limit of a punctual source  $a \rightarrow 0$, 
one gets $\Psi(x) = - i\sqrt{\Log (1 -l^2/x^2)}$, 
where $x>0$  is assumed for convenience.  
The concentration and velocity fields can be written explicitly  
by using the relation $\Log \zcom    = \ln |\zcom| + i \Arg \zcom$  for  a complex number~$\zcom$.  

The  resulting steady state is illustrated in Fig.~\ref{fig:steadystate}a. 
First, one can check that the flux of surfactant is constant from the source to the sink. 
Second, the concentration vanishes for $|x|>l$ whereas the velocity does not. 
This illustrates the non-local relation between concentration and velocity. 
Third, 
one can note the diverging slope of concentration and velocity profiles 
in the vicinity of sources and sinks. 
Finally, 
the concentration is clearly distinct  from the linear profile expected with pure diffusion. 
The one-dimensional steady source with sinks 
is another situations where Marangoni spreading is not diffusive-like. 

\paragraph*{With evaporation.}
Evaporation acts as a distributed sink and presumably ensures the existence of a steady state. 
Even with the simplest case of a Cauchy source, 
the integration of Eq.~\eqref{eq:steadysource} leads to an equation on $\Psi$ that is untractable. 
We can however look for an expansion of $\Psi$ at small and large~$x$. 
For small~$x$, assuming $-i \Psi(x)=c_0 + c_1 i x  + c_2 x^2$ 
with all coefficients real leads to an approximation that depends only on $c_0$.  
One finds 
$c_1= 1/a c_0 - \alpha$ and $c_2=(\alpha  a c_0+c_0^2-1)/(2 a^2 c_0^3)$. 
The coefficient $c_0$ is solution of a transcendantal equation that can only be solved numerically. 
In the vicinity of the source, 
the concentration profile is thus parabolic and the velocity profile is linear. 
For large~$x$, assuming a power series $\Psi(x) = \sum_{n=1}^\infty  c_n x^{-n}$ 
leads at lowest order to 
$\Psi(x) = 1/\alpha x - i a/\alpha x^2 + ...$ 
Far from the source, 
the velocity and concentration fields decay as $x^{-1}$ and  $x^{-2}$ respectively. 
Figure~\ref{fig:steadystate}(b) shows a solution of Eq.~\eqref{eq:steadysource}
obtained numerically, together with the approximations above. 
The latter are quite satisfactory, except for position around unity.   
Note finally that, consistent with the absence of steady solution when~$\alpha=0$, 
the limit of vanishing evaporation can not be taken.

\section{Riesz transform of a radial function}
\label{sec:appriesz}


Obtaining the Riesz transform is not straightforward, 
even with the assumption of radial symmetry. 
Accordingly, we give two expressions that we found useful in this purpose. 
From its definition in  Fourier space of Eq.~\eqref{eq:uqfromfq}, 
the Riesz transform may be recast as 
\begin{align}
 \riz{f}(r) = - \partial_r \left[   \fouop^{-1}_{q \rightarrow r} \left(  \frac{1}{q}  \fouop_{r \rightarrow q} (f(r))  \right)   \right], 
\end{align}
where $\fouop$ indicates a Fourier transform for a purely radial two-dimensional function: 
\begin{subequations}
\begin{align}
  \fouop_     {r \rightarrow q} f  & \eqdef           2 \pi \int_0^\infty f(r) J_0(q r) \,r\, \dif{r}, \\
  \fouop^{-1}_{q \rightarrow r} f  & \eqdef\frac{1}{2 \pi} \int_0^\infty f(q) J_0(q r) \,q\, \dif{q}, 
\end{align}
\end{subequations}
with $J_m$~the Bessel function of the first kind of order~$m$.
An equivalent expression can be given in terms of Hankel transform~\footnote{The expression holds provided all involved integrals are convergent.} 
\begin{align}
 \riz{f}(r) = \hanop^{-1}_{1, q \rightarrow r} \left[  \hanop_{0, r \rightarrow q} f(r) \right],   
\end{align}
where  $\hanop_\nu$ is the Hankel transform of order~$\nu$:
\begin{subequations}
\begin{align}
\hanop_{\nu, r \rightarrow q}      f  & \eqdef \int_0^\infty f(r) J_\nu(q r) \,r\, \dif{r}, \\
\hanop_{\nu, q \rightarrow r}^{-1} f  & \eqdef \int_0^\infty f(q) J_\nu(q r) \,q\, \dif{q}.
\end{align}
\end{subequations}

As a side remark, 
we note that the Riesz transform may be interpreted by analogy with electrostatic or gravitational field. 
If $f(r)$ is the radial density of charge on an infinitely thin disk, 
$\riz{f}(r)$ is the radial component of the electric field in the plane of the disk. 
Let us then assume that  identically charged particles are restricted to move in a plane 
and in a quiescent medium so that they have a fixed mobility, 
\ie{} a linear relationship between their velocity and the force acting on them. 
Then, if non-electrostatic interaction can be neglected, 
the evolution of the charge density is governed by Eq.~\eqref{eq:eq2Dinsolubleradial} 
and the process is analogous to  Marangoni spreading.

\begin{widetext}
\section{Long formulas}
\label{sec:appformula}

We give here some formulas that were too long to be included in the main body of the text in Sec.~\ref{holes}.

\paragraph*{The Cauchy hole:}
\begin{subequations}
\begin{align}
2 t \Psi(x,t) =&  x + i a   - i \cinf t                                                        
              -i \sqrt{a^2 -a (4 \am t - 2 \cinf t+2 i x)+(\cinf t-i x)^2},   \hspace*{-10cm} &                 \\
 f(x,t)       =& \frac{1}{2 t} \left( \Upsilon^{+} -a + \cinf t \right),                        
               & \Upsilon^\pm  &\eqdef \sqrt{\sqrt{ \xi^2 +  (2 (a + \cinf t) x)^2} \pm \xi } / \sqrt{2},   \\
 u(x,t)       =& \frac{1}{2 t} \left( x - \Upsilon^{-} \right),                                 
               &  \xi         &\eqdef a^2 + 2 a (-2 \am + \cinf) t + \cinf^2 t^2 - x^2.                           
\end{align}
\end{subequations}

\paragraph*{The circular hole:} 
\begin{subequations}
\begin{align}
 \Psi(x,t )     =& \frac{\am \left(\sqrt{-a^2+2 a \am t-(\cinf t-i x)^2}+i \cinf t-x\right)-i a \cinf}{a-2 \am t},   \hspace*{-8cm} & \\
 f(x,t)         =&       \frac{\am}{ 2 \am t - a} \left(  \Upsilon^{+} - a \cinf/\am +  \cinf t  \right),           
                 & \Upsilon^{\pm} &\eqdef  \sqrt{ \sqrt{\xi^2 +  (2 \cinf t x)^2} \pm \xi} /\sqrt{2},        \\ 
 u(x,t)         =&       \frac{\am}{ 2 \am t - a} \left( x - \Upsilon^{-} \right),                                     
                 & \xi            &\eqdef   a^2 - 2 a \am t + \cinf^2 t^2 - x^2.                                            
\end{align}
\end{subequations}

\end{widetext}

\bibliographystyle{bibgenym}

\end{document}